# Laser writing of scalable single colour centre in silicon carbide

Yu-Chen Chen[1*], Patrick S. Salter[2], Matthias Niethammer[1], Matthias Widmann[1], Florian Kaiser[1], Roland Nagy[1], Naoya Morioka[1], Charles Babin[1], Jürgen Erlekampf[3], Patrick Berwian[3], Martin Booth[2], Jörg Wrachtrup[1]

1. 3rd Institute of Physics, University of Stuttgart and Institute for Quantum Science and Technology IQST, Germany

2. Department of Engineering Science, University of Oxford, Parks Road, Oxford OX1 3PJ, UK

3. Fraunhofer IISB, D-91058 Erlangen, Germany

*Corresponding author: y.chen@pi3.uni-stuttgart.de

**Abstract:**

Single photon emitters in silicon carbide (SiC) are attracting attention as quantum photonic systems [1-2]. However, to achieve scalable devices it is essential to generate single photon emitters at desired locations on demand. Here we report the controlled creation of single silicon vacancy ($V_{Si}$) centres in 4H-SiC using laser writing without any post-annealing process. Due to the aberration correction in the writing apparatus and the non-annealing process, we generate single $V_{Si}$ centres with yields up to 30%, located within about 80 nm of the desired position in the transverse plane. We also investigated the photophysics of the laser writing $V_{Si}$ centres and conclude that there are about 16 photons involved in the laser writing $V_{Si}$ centres process. Our results represent a powerful tool in fabrication of single $V_{Si}$ centres in SiC for quantum technologies and provide further insights into laser writing defects in dielectric materials.

**Introduction**

Silicon carbide (SiC) is host to many promising colour centres with extremely high brightness [3-7] and long spin coherence time [8-11], showing favourable properties as quantum light sources and providing optical interfaces with electron and nuclear spins. In particular, silicon vacancy ($V_{Si}$) defects in SiC have shown to be useful for high-precision vector magnetometry [12,13], all-optical magnetometry [14], thermometry [15] and quantum photonics [16,17].

To provide scalable quantum technologies based on $V_{Si}$ centres in SiC, it will be necessary to integrate those systems into nanophotonics and electronics functional structures [8,16,17] with an accuracy of 10 nm - 1 µm. In this regard, focused Si ion and proton beams have been used to engineer $V_{Si}$ centres in SiC with high spatial accuracy [18,19]. However, those methods come at the expense of creating considerable residual lattice damage, which may reduce spin and optical coherence properties. Recently, the laser writing method was successfully implemented to generate single nitrogen vacancy centres in diamond with high positioning accuracy and near-unity yield, while maintaining excellent spin and optical coherence properties [20-22]. Moreover, laser writing method has recently successfully generated $V_{Si}$ centres ensembles in 4H-SiC at various depth [23].

In this work, we apply the laser writing method to fabricate single $V_{Si}$ centres in 4H polytype of SiC with high yield and good optical properties. Due to the combination of nonlinear laser writing process and appropriate aberration correction [20, 21], our method can generate defects at any depth in the crystal and with a spatial resolution beyond the optical diffraction limit. We also discuss the mechanism of the laser writing process.

**Main text**

Our host material is a 4H-SiC epitaxial layer that was grown along c-axis in an Epigress CVD reactor (VP508) at 1625 °C with nitrogen doping of $1 \times 10^{15}\ cm^{-3}$ and the a thickness of $62\ \mu m$ on an 4° off-axis 4H-SiC substrate. Then, the sample was oxidized in oxygen atmosphere at 1300 °C for 5.5 hours and subsequently annealed in Argon atmosphere at 1500 °C to minimize the carbon vacancy ($V_C$) concentration.

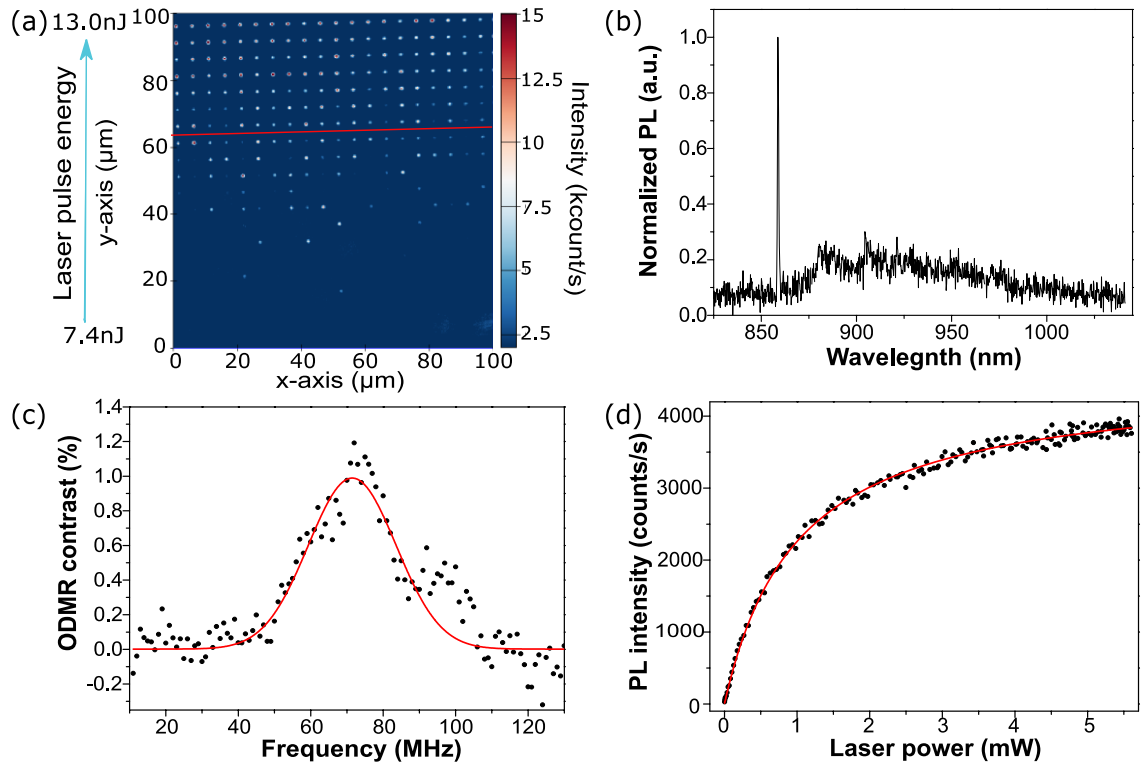

*Figure 1.* **Generation of $V_{Si}$ centres using laser writing.** (a) PL image of Array A immediately after laser processing. The laser pulse energy increases from the bottom to the top of the image and the range of the laser pulse energy used are provided on the left of the figure. The red line at the pulse energy of 12.6 nJ indicates the pulse energy threshold to generate colour centres in all 20 repeats. (b) Typical spectrum at 4.2 K of the features generated with laser pulse energy lower than 12.6 nJ, which indicates the generation of hexagonal lattice site $V_{Si}$ centres (V1'). (c) Typical ODMR signal of the laser written $V_{Si}$ centres ensemble at room temperature, revealing cubic lattice site $V_{Si}$ centres (V2). (d) Typical background-corrected excitation power dependent fluorescence of laser written single $V_{Si}$ centre. The linearly increasing background fluorescence was recorded by when being aligned for away from the emitter.

We characterized the sample via photoluminescence (PL) detection using a confocal microscope arrangement (see Supporting information). Our initial PL scans of the sample allowed us to identify large regions without 'native' $V_{Si}$ centres, which were subsequently chosen to conduct the laser writing experiments. Single writing pulses of wavelength 790 nm and duration 250 fs were delivered to each site in 40 x 20 square grids with a pitch of 5 µm at a depth of 20 µm. Along the vertical axis of the grids, the pulse energy, $E_p$, was varied from 6.7 nJ to 89.4 nJ to generate incremental degrees of damage to the lattice (see Supporting Information for more details). Along the horizontal axis, 20 identical pulses were delivered to perform statistics for each pulse energy. In this paper, this array is labelled as Array A. Additionally, we also generated a second array at a deeper depth of 40 µm with single writing pulses for each site to demonstrate the 3D ability of laser writing (see Figure S2 in Sporting Information). This deeper array will be called Array B.

Figure 1a shows a PL image of Array A immediately after laser writing without annealing or other treatments. Since the size of Array A is much larger than the scan range of our confocal microscope, we only show the PL image of the area exposed to the laser pulse energy from 7.4 to 15.6 nJ in Figure 1a (from bottom to top). At a threshold pulse energy of $E_p \geq 12.6 \ nJ$ fluorescence was observed at each site (red line in Figure 1a). As $E_p$ decreases from 12.6 nJ, the probability for defect generation among 20 repeats drops. Furthermore, no visible fluorescent sites are observed for $E_p \leq 8.2 \ nJ$. These findings indicate that the average number of colour centres' per site decrease as a function of laser pulse energy.

Figure 1b shows the typical emission spectrum of a laser written feature at 4.2 K. The spectrum shows a sharp peak at 858.7 nm with roughly 100 nm wide phonon side band, which corroborates the generation of hexagonal lattice site silicon carbide centres [24]. Note that the sharp peak at 858.7 nm corresponds to the optical transition between the ground state and the second excited state V1' of the $V_{Si}$ centre [25]. Since in our optical arrangement, we collect fluorescence almost parallel to the c-axis of 4H-SiC, we cannot observe emission from the first excite state V1, whose dipole orientation is parallel to the c-axis [24,25]. For the same reason, we do also not observe emission from cubic lattice site $V_{Si}$ centres (V2). They are even not seen in the spectrum of the heavy damage alignment markers (see Supporting Information), as shown in Figure S4 in the Supporting Information. At sites fabricated with high laser pulse energy, there may be many V2 $V_{Si}$ centres generated, but their emission is weak and covered by the V1' emission. However, due to the different zero-field splitting energies of V1 and V2 $V_{Si}$ centres, the optically detected magnetic resonance (ODMR) measurements can distinguish the emission of V2 from V1 $V_{Si}$ centres. The ODMR measurements were performed on the laser written $V_{Si}$ centres ensemble at room temperature. 6 out of 10 sites written with $E_p \geq 22.7 \ nJ$ show a clear ODMR signal at around 70 MHz at zero external field indicating the existence of V2 $V_{Si}$ centres by its characteristic zero field splitting, as shown in Figure 1c. The red line is the best fit to a Gaussian function, giving the peak position at 71.5 ± 0.41 MHz. The observed side peaks are due to a small inhomogeneous parasitic magnetic field from the experimental apparatus [13].

Figure 1d shows the excitation power dependent fluorescence intensity for a single hexagonal site $V_{Si}$ centre at room temperature. The intensity $I(P)$ can be described by the formula

$$I(P) = \frac{PI_{sat}}{P + P_{sat}}, \qquad (1)$$

where $I_{sat}$ is the saturation intensity and $P_{sat}$ is the saturation power of the emitter. The red line in Figure 1d is the best fit to Equation (1), giving $I_{sat} = 4526 \pm 17$ counts/s and $P_{sat} = 1.00 \pm 0.02$ mW. Note that $I_{sat}$ is 2 to 3 times lower than previously reported values, which is due to the deep location of the laser written $V_{Si}$ centres [8, 19, 26]. We expect similar PL intensity when the defects are created more shallow (few μm below surface).

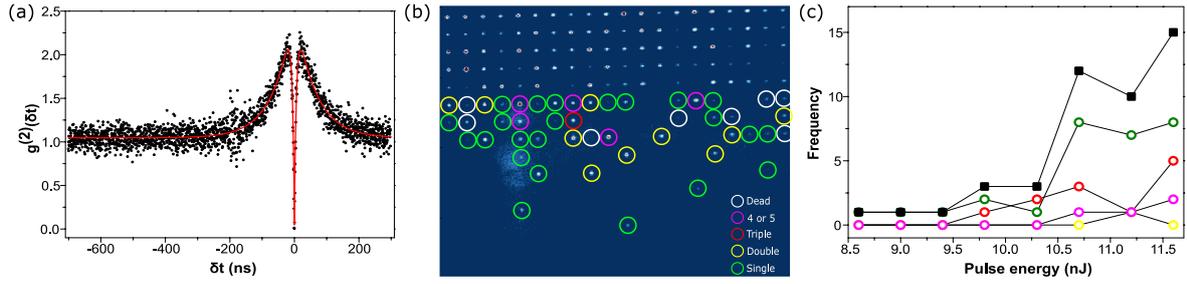

*Figure 2.* **Statistics of V$_{Si}$ centres generation using laser processing.** *(a) Typical $g^{(2)}(\delta t)$ histogram from single V$_{Si}$ centre after background removal at room temperature. (b) Map of the number of V$_{Si}$ centres generating at different sites in Array A. 'Dead' refers to the centres which photobleached during $g^{(2)}(\delta t)$ measurements. (c) Plot of the number of single (green), double (yellow), triple (red), and 4 or 5 (purple) V$_{Si}$ centres generated in each row of 20 repeats as a function of laser pulse energy measured before the objective lens in the writing apparatus. The total number generated per row is presented in black.*

Photon autocorrelation measurements (see Methods) were conducted for each site with $E_p \leq 11.6\ nJ$ in Array A to determine the number of V$_{Si}$ centres. Since the detected PL count rate is low (4-14 kcounts/s), the background fluorescence (about 1.0 kcounts/s) needs to be accounted for in the second order autocorrelation histogram $g^{(2)}(\delta t)$ (see Method for detail). Fig 2a shows a typical data set displaying a distinct photon anti-bunching at δt = 0 with $g^{(2)}(\delta t) = 0.040 \pm 0.003$, which stands as a clear characteristic of a single colour centre. In addition, for $|\delta t| > 16.5$ ns we observe a significant bunching behaviour, which is typical for a three-level like the one investigated here (see Method).

Figure 2b shows a spatial map of the V$_{Si}$ populations per laser writing site of Array A. In spots that have been created using lower pulse energies, all V$_{Si}$ centres are optically stable. At sites $E_p > 10.7\ nJ$, 2 to 3 laser written V$_{Si}$ centres among 20 repeats were photobleached after a few hours optical excitation and 1 to 2 V$_{Si}$ centres showed blinking behaviour. Combining these observations, we suggest that the optical instability may be attributed to the charge transfer from the V$_{Si}$ centres to other laser generated defects in the vicinity, such as carbon vacancies (V$_C$), C interstitial and Si interstitial defects. To our best knowledge, these phenomena were not observed in V$_{Si}$ centres generated by electron irradiation. A likely explanation is that, in laser writing, there is a higher chance of creating the V$_C$ centres, C interstitial and Si interstitial defects in the vicinity of the V$_{Si}$ centres, as the laser energy is focused into a small volume around the position where the V$_{Si}$ centre is generated. Oshima et al proposed that the post irradiation annealing may be able to remove unwanted defects and stabilize the V$_{Si}$ centres [27]. The optical stability of laser written V$_{Si}$ centres may be improved by adjusting the n-doping level [28]. Although some V$_{Si}$ centres are not optically stable, at 10.7 nJ, 6 out of 20 sites revealed an optically stable single V$_{Si}$, corresponding to a probability of 30%. Figure 2c shows the statistics of each row in Array A as a function of the laser writing pulse energy. Here, the number of V$_{Si}$ centres in photobleached sites is estimated from the initial PL count rate. The total number of V$_{Si}$ centres generated per row shows a systematic increase with the laser pulse energy. The probability of fabricating single V$_{Si}$ centres also increases with writing pulse energy and reaches its maximum at after a pulse energy of 10.7 nJ. The number of multiple V$_{Si}$ centres per row increase with pulse energy, as well.

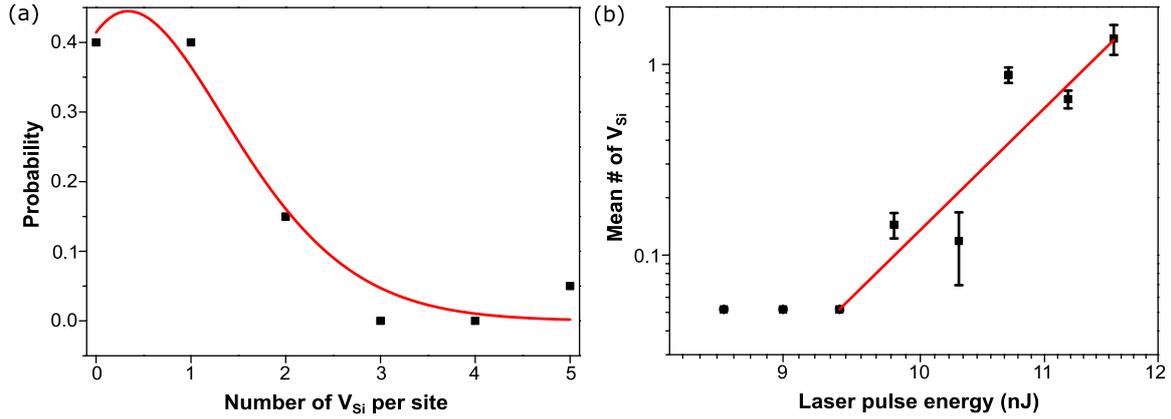

*Figure 3. **Nonlinearity of the laser writing process.** (a) Statistical distribution of the number of the $V_{Si}$ centres per laser writing site at laser pulse energy of 10.7 nJ. The probability value is the ratio between the number of site for each case of number of $V_{Si}$ per site and 20 repeats. (b) The average number of the $V_{Si}$ centres per laser writing site among 20 repeats as a function of laser pulse energy.*

In the following, we discuss the mechanism of laser writing of the $V_{Si}$ centres in SiC. First, we consider the statistics of the distribution of the population of $V_{Si}$ centres per site for 20 repeats for each writing pulse energy. Figure 3(a) shows the results of sites written with a pulse energy of 10.7 nJ, as an example. The red line in Figure 3(a) is a least-squares fit of the Poisson distribution function, $P_\lambda(k) = \frac{\lambda^k e^{-\lambda}}{k!}$. Here $\lambda$ is the average number of $V_{Si}$ centres per site and $k$ is the number of $V_{Si}$ centres generated per site. At writing pulse energy of 10.7 nJ, the Poisson distribution fit gives $\lambda = 0.88 \pm 0.08$. The mean value for generated $V_{Si}$ from the Poisson fits is plotted in Figure 3b as a function of writing pulse energy with the error bars representing the fitting error. Note that any laser generated $V_{Si}$ centres which may have been present but in the wrong charge state at the beginning or were photobleached during PL scan are not taken into account. Thus, our data analysis represents a lower limit case. In the region between 8.6 and 9.4 nJ, the average number of generated $V_{Si}$ centres remains constant, possibly because there are not enough repeats to sufficiently evaluate them statistically. For this reason, we exclude the first two data points in the following analysis. The average number of $V_{Si}$ centres per site exponentially increases with the writing pulse energy at $E_p > 9.4\ nJ$, which is similar to laser writing defect generation in diamond [29]. Lagomarsino et al in ref 28 demonstrated that the dominant free electrons generation mechanism at low writing pulse energy is multiphoton ionization (MPI). Thus, we think that the MPI can describe the mechanism of the laser writing $V_{Si}$ centres as the writing pulse energy is lower than 12 nJ (see Supporting Information). The photonionization rate of MPI can be expressed as

$$P(E) = \sigma_k E_p^k, \quad (2)$$

where $\sigma_k$ is the multiphoton absorption coefficient for the absorption of $k$ photons and $E_p$ is the writing pulse energy [30]. The red line in Figure 3b is the least-squares fit of Equation (2), giving $k = 15.5 \pm 1.2$. In other words, each photon in the 790 nm laser pulse possesses an energy of 1.57 eV, implying that energy is transferred to the SiC lattice in packets of 24.33 ± 1.88 eV. The displacement threshold energy for Si atom in SiC is approximately 25 eV [31], which falls within the error bar of our result. Note that the 15.5 photon absorption obtained here is the nonlinearity of vacancy generation instead of the free electron generation in ref 29. To date, it remains still unclear how the free electrons convert to defects for ultrafast laser pulse writing in wide bandgap materials. The nonlinearity analysis merely provides an insight into certain aspects of the mechanism for defects generation in SiC.

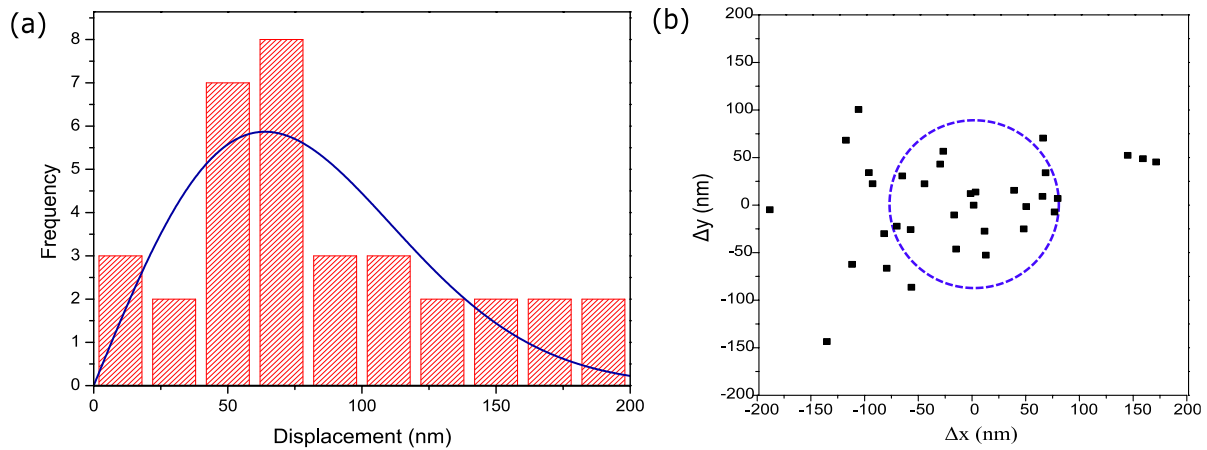

Figure 4. **Positioning accuracy of the $V_{Si}$ centres generation using laser processing.** (a) Histogram of the displacement in the image plane for the $V_{Si}$ centres. The data are fitted with a Rayleigh distribution function. (b) Scatter plot of the fitted positions of the $V_{Si}$ centres relative to the theoretical grid points in the image plane. The dashed circle represents the mean displacement obtained from fitting in (a).

For applications, it is important that the $V_{Si}$ centres can be accurately positioned in 3D in order to integrate them into other device architectures. The precision with which we can position $V_{Si}$ centres in the image plane was obtained by fitting a 2D Gaussian surface to the measured intensity distribution of each laser written $V_{Si}$ centre. The theoretical grid points were determined by conducting a least-squares fit of a regular square grid with 5 µm grid length to laser written $V_{Si}$ centres' positions using fitting parameters of displacement in x and y. The obtained displacements between the $V_{Si}$ centres and the theoretical grid points are plotted as a histogram in Figure 4a. The blue line is the best fit to a Rayleigh distribution, giving a mean displacement of 80 ± 9 nm. Figure 4b shows the spatial distribution of the $V_{Si}$ centres' positions relative to their theoretical grid points and the dashed circle is the mean displacement. The mean displacement is a factor of 3.7 smaller than the diffraction limited width of the focal spot of the writing laser (see Supporting Information), which underlines the great potential of the laser writing method to precisely generate on-demand defects in photonic structures.

The nonlinearity of the MPI mechanism is expected to result in vacancy creation within a region smaller than the focal volume estimated by the field intensity $I(r,z)$ (Equation S1 in Supporting Information). According to MPI model, the spatial distribution function for vacancy generation via a y-photon absorption process is dependent on $I^y(r,z)$ and the corresponding full width at half maximum (FWHM) in radial and axial directions are given by

$$d_y = \omega_0 \sqrt{\frac{2ln2}{y}}, L_y = 2z_R\sqrt{2^{\frac{1}{y}}-1}. \qquad (3)$$

In the previous section, we have demonstrated that 15.5 photons are involved in the vacancy generation process. As a result, the $d_{15.5}$ and $L_{15.5}$ are corresponding to 88.8 and 364.4 nm, respectively. The estimated radial FWHM matches well with the measured mean displacement. We note that the positioning accuracy could be better than the estimated radial FWHM. The probability of vacancy creation is related to the writing pulse energy, and there is a threshold energy required before there is a significant probability of vacancy generation. When the writing pulse energy is only slightly above the threshold, the volume with measurable probability of vacancy generation could be smaller than the estimated volume from Equation 3. Chen et al recently demonstrated that the radial positioning accuracy of laser written single NV[-] centres in diamond is about 40 nm [21], which is half the measured positioning accuracy of $V_{Si}$ centres here. This may be due to a little spherical aberration in our PL image. In addition, many $V_{Si}$ centres created by relatively high writing pulse energy were included in the positioning accuracy analysis in this work.

**Summary**

In summary, we have demonstrated femtosecond laser writing of single $V_{Si}$ centres into SiC at predefined positions. The yield of optically stable single $V_{Si}$ centres is up to 30%. In diamond, some form of annealing is always required to generate NV[-] centres [20-22]. On the contrary, single $V_{Si}$ centres in SiC can be created with just a single pulse and no annealing. We note that additional annealing may be used to remove unwanted laser generated defects in order to improve the stability of the $V_{Si}$ centres and further improve the yield. The $80 \pm 9$ nm positioning accuracy in the image plane was achieved, which is sufficient for coupling $V_{Si}$ centres to optical structures such as multimode waveguides, whispering gallery resonators [32], solid immersion lenses [8], or nanopillars [16]. The positioning accuracy may be limited by the relatively high writing pulse energy rage, aberrations of detection apparatus and translational uncertainty of the stage control in the laser writing apparatus. The positioning accuracy of generating $V_{Si}$ centres can be possibly improved down to 40 nm, which has been already demonstrated in diamond [21], providing sufficient resolution to couple to nanophotonic structures [17]. The targeted laser writing $V_{Si}$ centres technique presented here can be applied on SiC with different polytypes, such as 3C and 6H. Since the count rate of $V_{Si}$ centres fluorescence is low, it is necessary to couple the $V_{Si}$ centres to photonic structure to measure the spin coherence time and line width of ZPL. In the future, we will laser write the $V_{Si}$ centres into solid immersion lenses or nanocavities to investigate the spin and optical coherence properties. Also, direct detection during laser writing might be used to allow the on-demand vacancy creation with instant success feedback. In addition, the laser writing method and subsequent annealing may generate other colour centres in SiC, such as divacancy [9,11], carbon antisite-vacancy pairs $(V_{Si}V_C)^0$ [6] and nitrogen vacancy $(N_CV_{Si})^-$ [33].

**Method**

Photon autocorrelation measurements and data fitting

Photon autocorrelation datasets were measured using the Hanbury Brown and Twiss method, using continuous wave 730 nm excitation at a power of 1.2 mW. A 850 nm long-pass filter was used for spectrally filtering. Background signals were measured by recording the fluorescence signal from the bulk SiC at a position off-focus from the laser writing site, used to calculate the baseline for the autocorrelation dataset according to the function

$$a = \frac{S}{S+B}$$

where S and B are the photon count rates for the colour centres and the background fluorescence respectively. The background baseline of the raw datasets, $g^{(2)}_{raw}(\delta t)$, is corrected according to the formula

$$g^{(2)}(\delta t) = \frac{g^{(2)}_{raw}(\delta t) - (1-a^2)}{a^2}$$

The dataset with background correction in Figure 2a were fitted by a standard function for the photon autocorrelation of a three-level system:

$$g^{(2)}(\delta t) = 1 - c\exp\left(\frac{-|\delta t|}{\tau_1}\right) + (c-1)\exp\left(\frac{-|\delta t|}{\tau_2}\right)$$

where $c$, $\tau_1$ and $\tau_2$ are related to the interlevel rate constants. The solid red line is a least-squares fit of this equation to the experimental data.

**Acknowledgements**

The work was supported by grants from the Baden-Württemberg Stiftung Programm: Internationale Spitzenforschung, the ERC SMel and the BMBF BRAINQSENS.

**Author contribution**

Y.-C. C. carried out the PL, spectrum and HBT measurements with assistance from M. N., M. W., F. K., R. N., N. M. and C. B. and coordinated the work. P. S. S. conducted the laser writing. J. E. grew the SiC sample with supervision from P. B. Y.-C. C., J. W. and M. B. conceived and oversaw the project. All authors contributed to writing the manuscript.

**Competing financial interests**

The authors declare no competing financial interests.

**Supporting information**

Laser writing processing

The optical layout for the laser writing is shown in Figure S1. The laser processing was performed using a regeneratively amplified Ti:Sapphire laser (Spectra Physics Solstice) at a wavelength of 790 nm and a 1 kHz pulse repetition rate. The laser beam was expanded onto a liquid crystal phase-only spatial light modulator (SLM) (Hamamatsu X10468-02), which was imaged in a 4f configuration onto the back aperture of a 60× 1.4 NA Olympus PlanApo oil immersion objective. The SiC sample was mounted on precision translation stages (Aerotech x-y: ABL 10100; z: ANT95-3-V) providing three dimensional control. Prior to the objective the laser pulse was linearly polarised and had a duration which was measured to be 250 fs using an intensity autocorrelator (APE Pulsecheck). The pulse duration at focus will be slightly increased due to dispersion in the objective lens. To optimize the aberration correction, the phase pattern displayed on the SLM was adjusted to minimize the pulse energy needed to produce visible fluorescence at test processing positions of similar depth in the sample.

The point spread function of the processing laser is that of a focused Gaussian beam

$$I(r,z) = I_0 \cdot \frac{1}{1+\left(\frac{z}{z_R}\right)^2} \cdot e^{-\frac{2r^2}{w_z^2}} \quad (S1)$$

where $w_z = w_0\sqrt{1+\left(\frac{z}{z_R}\right)^2}$ is the beam width at axial displacement $z$, $w_0$ is the beam waist and $z_R$ is the Rayleigh range. Based on the numerical aperture of the objective lens, we estimate that $w_0 = 297$ nm and $z_R = 852$ nm.

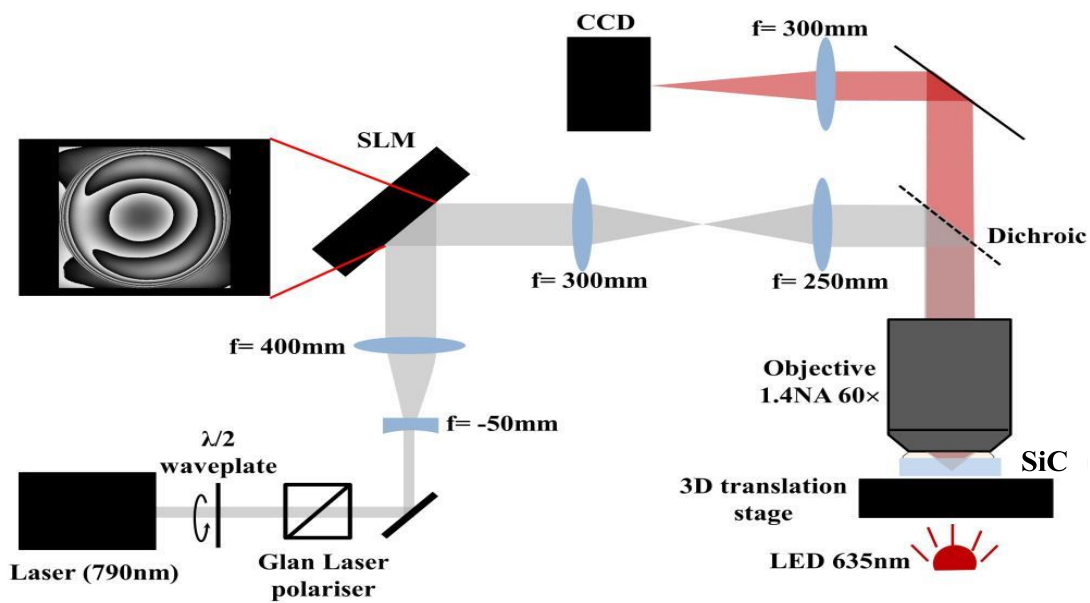

Figure S1. Aberration-corrected laser writing. (a) Schematic of laser processing apparatus. The phase pattern displayed on the SLM compensated optical aberration in the system and those introduced by focussing into the SiC sample.

Laser written array at depth of 40 μm

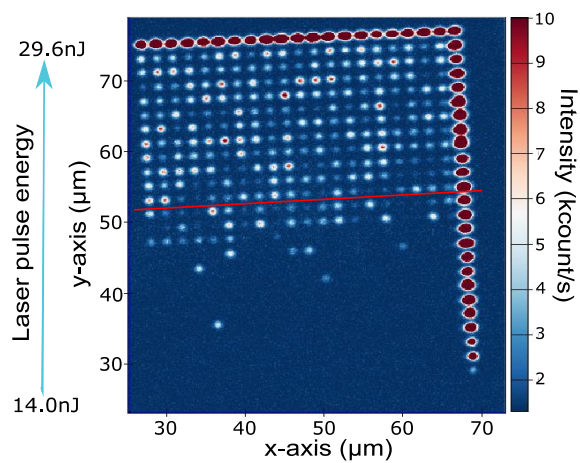

Figure S2. PL image if the Array B at depth of 40 μm below surface immediately after laser processing. The range of the laser pulse energy is provided on the left of the image. The red line at pulse energy of 22.7 nJ indicates the pulse energy to generate colour centres in all 20 repeats.

Array B was generated with single laser pulses for each site in 26 x 20 square grids with a pitch of 2 μm at a depth of 40 μm. Along one axis of the grids, the pulse energy, $E_p$, was varied from 29.6 nJ to 14.0 nJ to generate incremental degrees of damage to the lattice (see supporting information for detail). Along the other axis, 20 identical pulses were delivered to investigate statistics of the results for each pulse energy. The first column on right hand side of the array was created by 10 laser pulses of each energy and the first row on the top of the array was created by 5 laser pulses with laser pulse energy of 29.6nJ. These features generated by multiple pulses serve as markers to locate the array. The heavy damage markers were also generated around Array A. Visible fluorescence was produced at every laser writing sites when they had been exposed to pulses with $E_p \geq 22.7$ nJ (red line in Figure 2S).

Writing pulse energy

The laser pulse energy was precisely controlled using a rotatable half waveplate before a Glan-Taylor polariser. 40 pulse energies used for fabrication of the Array A and Array B are shown in table S1 and S2, respectively. The pulse energies were measured immediately before the objective. The pulse energies after the objective lens can be estimated by multiplying a factor of 0.7 due to the transmission loss of the objective lens, based on the manufacturer's specifications. For the pulse energy at the focus, one must also take into account Fresnel reflection at the SiC interface which can be significant due to the large angles involved.

| Laser pulse energy before objective (nJ) | Laser pulse energy after objective (nJ) |
|---|---|
| 89.4 | 62.58 |
| 77.3 | 54.11 |
| 66.1 | 46.27 |
| 55.6 | 38.92 |
| 46.1 | 32.27 |
| 42.5 | 29.75 |
| 39.1 | 27.37 |
| 35.8 | 25.06 |
| 32.6 | 22.82 |
| 29.6 | 20.72 |
| 26.7 | 18.69 |
| 24.0 | 16.8 |
| 21.4 | 14.98 |
| 19.0 | 13.3 |
| 18.4 | 12.88 |
| 17.8 | 12.46 |
| 17.3 | 12.11 |
| 16.7 | 11.69 |
| 16.1 | 11.27 |
| 15.6 | 10.92 |
| 15.1 | 10.57 |
| 14.6 | 10.22 |
| 14.0 | 9.8 |
| 13.5 | 9.45 |
| 13.0 | 9.1 |
| 12.6 | 8.82 |
| 12.1 | 8.47 |
| 11.6 | 8.12 |
| 11.2 | 7.84 |
| 10.7 | 7.49 |
| 10.3 | 7.21 |
| 9.8 | 6.86 |
| 9.4 | 6.58 |
| 9.0 | 6.3 |
| 8.6 | 6.02 |
| 8.2 | 5.74 |
| 7.8 | 5.46 |
| 7.4 | 5.18 |
| 7.1 | 4.97 |
| 6.7 | 4.69 |

Table S1 Laser pulses energies used for the writing of defects in Array A.

| Laser pulse energy before objective (nJ) | Laser pulse energy after objective (nJ) |
| --- | --- |
| 29.6 | 20.72 |
| 28.9 | 20.23 |
| 28.1 | 19.67 |
| 27.4 | 19.18 |
| 26.7 | 18.69 |
| 26.0 | 18.2 |
| 25.4 | 17.78 |
| 24.7 | 17.29 |
| 24.0 | 16.8 |
| 23.3 | 16.31 |
| 22.7 | 15.89 |
| 22.1 | 15.47 |
| 21.4 | 14.98 |
| 20.8 | 14.56 |
| 20.2 | 14.14 |
| 19.6 | 13.72 |
| 19.0 | 13.3 |
| 18.4 | 12.88 |
| 17.8 | 12.46 |
| 17.3 | 12.11 |
| 16.7 | 11.69 |
| 16.1 | 11.27 |
| 15.6 | 10.92 |
| 15.1 | 10.57 |
| 14.6 | 10.22 |
| 14.0 | 9.8 |

Table S2 Laser pulses energies used for the writing of defects in Array B.

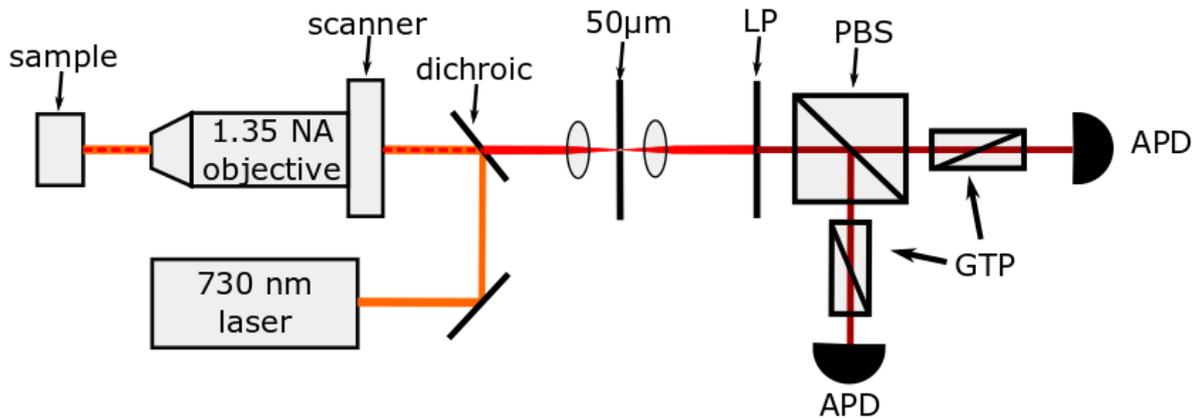

Figure S3. Experimental setup. Orange line represents the 730 nm excitation beam, the red line shows the fluorescence, dark red is the filtered fluorescence. LP indicates 850 nm long pass filter, PBS is polarizing beam-splitter, GTP is Glen-Taylor polarizer and APD is single photon Avalanche photon detector.

Optical characterization of the sample is carried out by a home-built confocal microscope depicted in Figure 2S. A red diode laser (Thorlabs HL7302MG, emission peak wavelength 730 nm, 40 mW) was focused via an oil objective (Olympus, UPLSAP 60×, NA 1.35) onto the sample to excite the silicon vacancies. The objective is mounted on a piezo-electric stage with a travel range in x-y-z of (100 × 100 × 10) µm, respectively. Fluorescence from the emitters is collected by the same objective and passes through the dichroic mirror. Then, the fluorescence is focused into a 50 µm pinhole to form confocal and spectrally filtered by an 850 nm long pass. The polarizing beam splitter (PBS), combined with two avalanche photon detectors, form a Hanbury-Brown and Twiss (HBT) setup. The addition Glen-Taylor calcite polarizers (GTP, Thorlabs GT10-B) are placed to suppress flash-back light from the APDs (Perkin Elmer SPCM-AQRH-15), which is caused by the avalanche of charge carriers accompanied by light emission.

Electron spins were manipulated by a radio-frequency electromagnetic field generated by a signal generator (ROHDE&SCHWARZ, SMIQ03B) and subsequently amplified by a broadband amplifier (Minicircuits ZHL-42W). RF irradiation was delivered to sample via a 20 µm thick copper wire, which is placed close to the tested defects (typically ~30 µm). For other details, see refs [1,2].

The low temperature measurements were conducted at a cryogenic temperature 4.2 K in a Montana Instruments Cryostation. The excitation was performed with a 785 nm diode laser. The fluorescence is filter by an 850 long pass filter and detected by a near infrared enhanced single photon avalanche photodiode (Excelitas, SPCM-AQRH-W5). PL spectra was recorded using a Czerny-Turner type spectrometer (Acton, SpectraPro300i, grating: 300 g mm$^{-1}$), combined with a back-illuminated CCD camera (Princeton Instruments, LN/CCD1340/400 EHRB/I).

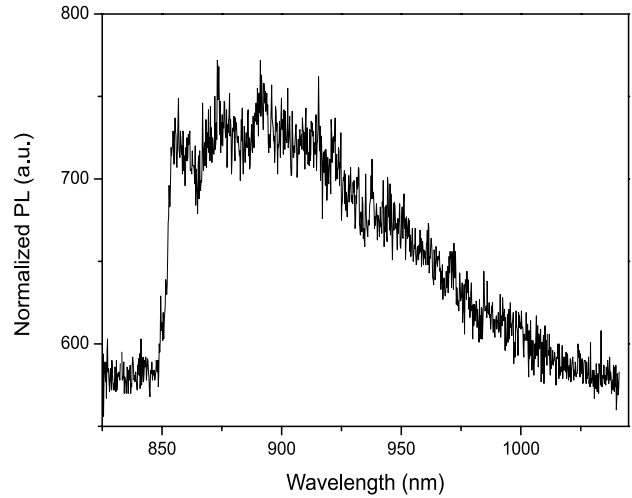

Figure S4. Typical spectrum of the marker generated by multiple laser pulses at 4.2 K.

Vacancy generation mechanism

Here, we discuss the mechanism of laser writing vacancies in dielectric materials. When a transparent dielectric material is illuminated by intense femtosecond laser, a large amount of excited electrons may be generated [3]. Relaxation channels of these free electrons in wide bandgap materials may produce intrinsic defects, resulting in photoinduced damages. For transparent materials, there is no linear absorption of the incident laser light to excite electrons due to the bandgap greater than photon energy. Therefore, a nonlinear mechanism must be responsible for the generation of free electrons creation [3, 4]. There two possible models have been proposed to describe the mechanism of free electron creation, which are tunnelling ionization (Zener breakdown) and multiple ionization (MPI). The details of these two model can be found in the ref. 3 and 4. The dominant mechanism for free electron generation is typically delineated by the Keldysh parameter [3]

$$\gamma = \frac{\omega}{e}\sqrt{\frac{mcn\varepsilon_0 E_g}{I}}$$

where $\omega$ and $I$ are the laser frequency and intensity, while $m, c, n, \varepsilon_0$ and $E_g$ are the electron effective mass, speed of light in vacuum, linear refractive index, permittivity of free space and the direct bandgap respectively. MPI is dominant over tunnelling breakdown for $\gamma > 1$ corresponding to

$$I < \frac{mcn\varepsilon_0 E_g \omega^2}{e^2}.$$

For the parameters used here ($m = 0.42 m_e$ [5], $n = 2.59, E_g = 3.23\ eV$ and $\omega = 2.4 \times 10^{15} s^{-1}$) assuming a pulse duration of 300 fs and a beam waist of 350 nm, the equation above predicts that MPI will dominate for pulse energies below about 12.27 nJ.